\documentstyle[prl,aps,preprint,epsf,tighten]{revtex}

\begin{document}
\setcounter{page}{0}
\def\footnoterule{\kern-3pt \hrule width\hsize \kern3pt}
\tighten
\title{Magnetohydrodynamics of the Early Universe and the Evolution of 
Primordial Magnetic Fields\thanks
{This work is supported in part by funds provided by the U.S.
Department of Energy (D.O.E.) under cooperative 
research agreement \#DF-FC02-94ER40818.}}
\author{D.T.~Son\footnote{Email address: {\tt son@ctp.mit.edu}}}
\address{Center for Theoretical Physics \\
Laboratory for Nuclear Science \\
and Department of Physics \\
Massachusetts Institute of Technology \\
Cambridge, Massachusetts 02139 \\
{~}}
\date{MIT-CTP-2724,~ hep-ph/9803412. March 1998, revised July 1998}

\maketitle

\thispagestyle{empty}

\begin{abstract}

We show that the decaying magnetohydrodynamic turbulence leads to a more
rapid growth of the correlation length of a primordial magnetic field than
that caused by the expansion of the Universe.  As an example, we consider
the magnetic fields created during the electroweak phase transition.  The
expansion of the universe alone would yield a correlation length at the
present epoch of 1 AU, whereas we find that the correlation length is
likely of order 100 AU, and cannot possibly be longer than $10^4$ AU for
non-helical fields.  If the primordial field is strongly helical, the
correlation length can be much larger, but we show that even in this case
it cannot exceed 100 pc.  All these estimates make it hard to believe that
the observed galactic magnetic fields can result from the amplification of
seed fields generated at the electroweak phase transition by the standard
galactic dynamo.

\end{abstract}

\vspace*{\fill}
%\begin{center}
%Submitted to: {\it Physics Letters B}
%\end{center}


\section{Introduction}

Recently, considerable interest has been focused on the possibility that a
primordial magnetic field may have been created at some early stage of the
evolution of the Universe.  While the existence of a weak widespread
extragalactic chaotic magnetic field cannot be ruled out, almost the only
place where such a primordial field may have left an observable imprint is
in galaxies, many of which possess microgauss, kiloparsec scale fields
that are thought to be the result of dynamo amplification of a weak seed
field \cite{Kronberg}.  The idea that the seed for the galactic dynamo may
be the field created in the very early Universe has inspired a number of
works.  For example, such fields may have appeared at the electroweak
phase transition \cite{BaymMcLerran}. There are numerous other proposals
involving physics at various scales (for a brief overview and further
references, see \cite{Olesen-rev}). 

One feature shared by most particle-physics scenarios is the smallness of
the correlation length of the magnetic field which results.  Indeed, at
the moment of creation, the correlation length is limited by the horizon
radius.  By pushing this moment to a very early stage in the history of
the Universe, one makes the correlation length much smaller than it would
be if the magnetic fields were created more recently, say, during
proto-galactic contraction \cite{Ostriker}.  For example, the fields
generated at the electroweak phase transition, when the horizon radius is
about 1 cm, would have a correlation length of at most about $10^{15}$ cm
in the present Universe.  Contraction of proto-galaxies is likely to
reduce this length scale by 2 orders of magnitude, which gives $10^{13}$
cm, or 1 AU, as the characteristic length scale of the seed field which
the galactic dynamo is supposed to amplify.  Realistically, this length
scale is likely much smaller, since the correlation length at creation is
typically much less than the horizon size\footnote{A magnetic field
created during or before inflation \cite{TurnerWidrow} may have a large
correlation length, however the proposed inflationary models that could
possibly generate a large-scale magnetic field may seem rather contrived
\cite{Ratra}.  We will not consider inflation-induced magnetic fields in
this paper, but one should keep in mind this alternative.}. 

Because any primordial magnetic fields generated at early times (via
particle physics occurring at, say, the electroweak scale) have such small
correlation lengths, these fields are not attractive candidates to play
the role of seed fields for the galactic dynamo, unless the correlation
length is somehow increased. The best developed theory of magnetic
amplification in galaxies, the mean-field dynamo \cite{Ruzmaikin},
operates under the assumption that the magnetic field is smooth on the
scale of turbulent motion of the interstellar gas, which is of order 100
pc.  One may attempt to apply the mean-field dynamo theory for large-scale
Fourier components of the chaotic magnetic field, neglecting the
small-scale ones, however it is not obviously the correct procedure.  That
the magnetic field at very small scales exponentiates rapidly is a well
known fact that poses a serious problem for the mean-field dynamo theory
\cite{KurlsrudAnderson}.  In the situation when the seed field itself
resides at small scales, this problem is likely to become more severe.

In this paper, we will not address the question of how a seed field is
created, whether at the electroweak phase transition or during some other
early epoch.  Our goal is to investigate the possibility that
magnetohydrodynamic (MHD) effects can lead to a substantial increase in
the length scale of the magnetic field at the present epoch.  We show that
decaying MHD turbulence typically leads to a faster growth of the magnetic
correlation length than one would expect from the expansion of the
Universe alone.  We estimate that in the case of magnetic fields generated
at the electroweak phase transition, the enhancement factor is $10^2$, and
the correlation length may reach 100 AU.  If the primordial field has a
large Chern-Simons number, the enhancement factor may be much larger, but
the correlation length cannot possibly exceed 100 pc today if the magnetic
field comes from the electroweak epoch.  Thus, although the
magnetohydrodynamic effects we consider are of some help, they cannot
increase the correlation length enough to make the electroweak generation
of a primordial seed field a viable option.  We also consider generating
the seed field at the QCD phase transition.  This may be a possibility,
but is only viable if the bubble separation at the phase transition is
very large.

This paper is organized as follows.  In Sec.\ \ref{sec:MHDeq} we write the
basic equation governing the MHD of the Universe.  Sec.\ \ref{sec:non-hel}
is devoted to the decay of non-helical MHD turbulence, whereas Sec.\
\ref{sec:hel} describes the scaling laws of the decay of helical
turbulence.  Sec.\ \ref{sec:concl} contains concluding remarks.  The
Appendix contains details about the EDQNM approximation used in our
numerical simulation.

\section{The MHD equations}
\label{sec:MHDeq}

To introduce notation, let us first write down the MHD equations for the
early Universe.  The MHD description applies since at interesting length
scales the Universe is a good conductor.  When the Universe expansion can
be neglected (that is, when the evolution occurs on a time scale much
smaller than the Hubble time), the MHD equations for a relativistic fluid
have the form (for a derivation see, e.g., Ref.\ \cite{Brandenburg}),
\begin{eqnarray}
  {\partial{\bf v}\over\partial t}+
  ({\bf v}\cdot\mbox{\boldmath$\nabla$}){\bf v} & = &
  -{1\over\rho+p}\mbox{\boldmath$\nabla$} \left(p+{B^2\over2}\right)+ 
  {1\over\rho+p}({\bf B}\cdot\mbox{\boldmath$\nabla$}){\bf B}+
  \nu\nabla^2{\bf v} \nonumber\\
  {\partial{\bf B}\over\partial t} & = &
  \mbox{\boldmath$\nabla$}\times({\bf v}\times{\bf B})+
  \eta\nabla^2{\bf B}
  \label{MHD} 
\end{eqnarray}
where ${\bf v}$ is the velocity of bulk fluid motion, ${\bf B}$ is the
magnetic field, $\rho$ and $p$ are the density and pressure of the fluid,
respectively.  We assume that the bulk flow velocity is non-relativistic,
$v\ll1$, although individual particles of the fluid move with the speed of
light.  The pressure can be eliminated from the equations by using the
incompressibility condition, $\mbox{\boldmath$\nabla$}\cdot{\bf
v}=0$\footnote{The fluid can be considered incompressible if the Mach
number, i.e. the ratio between $v$ and the sound speed, is much less than
1.  Since for ultra-relativistic fluids the speed of sound is
$1/\sqrt{3}$, this condition is essentially equivalent to $v\ll1$.}. The
kinematic viscosity $\nu$ and the resistivity $\eta$, both having the
dimension of length in units in which $\hbar=c=1$, are determined by
microscopic physics.  One can define the kinematic and magnetic Reynolds
numbers,
\[
  \text{Re}={vl\over\nu},\qquad \text{Re}_{\text{M}}={vl\over\eta}
\]
where $v$ and $l$ are the typical velocity and typical length scale of the
fluid motion.  As a rule, in the early Universe the magnetic Reynolds
number is much larger than the kinematic one (i.e.\ the magnetic Prandtl
number, defined as $\nu/\eta$, is large.)  Hydrodynamic turbulence occurs
when $\text{Re}\gg1$.  For example, at $T=100$ GeV, $\nu \sim (\alpha T
\log{1\over\alpha})^{-1}$, and $\eta \sim {\alpha\over T}\log
{1\over\alpha}$.  Taking $v\sim 10^{-1}$, $l\sim 10^{-2}a_H$, the two
Reynolds numbers are $\text{Re}\sim10^{11}$ and
$\text{Re}_{\text{M}}\sim10^{17}$, i.e. very large. 

Let us turn to the case of the expanding Universe.  In the
radiation-dominated epoch, neglecting the slow change of the number of
degrees of freedom with temperature, the MHD equations are of almost the
same form as Eqs.\ (\ref{MHD}) \cite{Brandenburg},
\begin{eqnarray}
  {\partial{\bf v}\over\partial \tau}+
  ({\bf v}\cdot\mbox{\boldmath$\nabla$}){\bf v} & = &
  -\mbox{\boldmath$\nabla$}\left(\tilde{p}+{\tilde{B}^2\over2}\right)+ 
  (\tilde{{\bf B}}\cdot\mbox{\boldmath$\nabla$})\tilde{{\bf B}}
  +\tilde{\nu}\Delta{\bf v} \nonumber\\
  {\partial\tilde{{\bf B}}\over\partial \tau} & = &
  \mbox{\boldmath$\nabla$}\times({\bf v}\times\tilde{{\bf B}}) + 
  \tilde{\eta}\Delta\tilde{{\bf B}}
  \label{MHD1}
\end{eqnarray}
where $\tau=\int\!dt\,a^{-1}(t)$ is the conformal time, $\tilde{{\bf
B}}={\bf B}/\sqrt{\rho+p}$ is defined in such a way that the conservation
of the magnetic flux during the expansion corresponds to the constancy of
$\tilde{\bf B}$.  The spatial derivative in Eq.\ (\ref{MHD1}) 
$\mbox{\boldmath$\nabla$}$ is taken with respect to the comoving
coordinates.  We have also defined $\tilde{\nu}=a^{-1}\nu$ and
$\tilde{\eta}=a^{-1}\eta$, which are functions of time not only because of
the factor $a^{-1}$ but also due to the temperature (and hence time) 
dependence of the viscosity and conductivity of the Universe.  The formal
coincidence of the MHD equations in the radiation-dominated Universe and
in the non-expanding Universe allows us to discuss the evolution of the
magnetic field in both cases from a common point of view.

\section{Decay of non-helical MHD turbulence}

\label{sec:non-hel}

For definiteness, let us discuss the magnetic field generated during the
electroweak phase transition, assuming the transition is first order
\cite{BaymMcLerran,Shaposhnikov}.  It has been suggested in that after the
phase transition, the cosmic fluid is in a state of turbulent motion, as
parts of the fluid moving in front of different expanding bubbles collide
\cite{BaymMcLerran}.  An important point that has not been sufficiently
emphasized in the literature is that until either another first-order
phase transition happens, or gravitational instability begins to act, no
additional stirring takes place and the turbulence will decay freely by
the turbulent energy dissipation (see below.)  As a consequence,
turbulence does not continue forever, but should terminate at some point.
Therefore, to establish the growth law of the magnetic correlation length,
one has to study the free decay of MHD turbulence, and its termination. 

Immediately after the phase transition, the characteristic velocity $v_0$
of the turbulence is that of the bubble walls.  The correlation length of
fluid velocity is the bubble separation $l_0$.  We will also assume the
presence of an initial chaotic magnetic field that has the same
correlation length as the velocity field and the same energy density as
the turbulent bulk fluid motion.  We rely here on the ``equipartition''
hypothesis, which states that any small seed magnetic field (which may
have been created during the phase transition \cite{BaymMcLerran} or come
from the hypercharge field created at some earlier epoch
\cite{Shaposhnikov}) will be enhanced by turbulent convection until the
magnetic energy is equal to the kinetic energy of the bulk fluid motion,
$\tilde{B}_0^2\sim v_0^2$.  Whether equipartition hypothesis is valid is a
controversial issue.  It is known that an arbitrarily weak magnetic field
is exponentially enhanced by turbulence when $\text{Re}_{\text{M}}$ is
larger than some critical value (of order 100, which is easily satisfied
in cosmological conditions), but it is unclear what is the saturation
level of the magnetic energy. However, it is very unlikely that we
underestimate the magnetic field by relying on equipartition.  As even
this magnetic field will prove insufficient, we will not worry whether we
are making an overestimate.  (The values of $v_0$ and $B_0$ mentioned
above correspond to the integral scale, i.e.\ the largest scale of
turbulence, which dominates the energy.) 

We will also presume for the present that the mean density of the
Chern-Simons number is vanishing.  This is the case if the initial
condition is parity symmetric in the statistical sense.  We discuss the
case of nonzero Chern-Simons number, and why it is different, in a later
section of this paper. 

\subsection{Non-expanding Universe: scaling laws}

Let us first consider the simpler case, in which the expansion of the
Universe can be neglected.  This corresponds to the first step of the
evolution, when the time passed from the creation of the magnetic field is
smaller than the Hubble time.  We will first make a simple argument of how
the correlation length grows with time, and then discuss the assumptions
behind this argument.  The mechanism for the growth is {\em selective
decay:} modes with larger wavenumbers decay faster than those whose
wavenumbers are small. Thus, at increasing time only modes with smaller
and smaller wavenumbers survive.  Let us assume that at some time $t$ only
modes with length scale larger than $l$ survive.  The modes that are about
to decay, i.e. those at the scale $l$, dominate the kinetic and magnetic
energies.  Denote the typical value of $v$ and $B$ of these modes as $v_l$
and $B_l$.  From the equipartition hypothesis, one has $\tilde{B}_l\sim
v_l$.  Let us find the dependence of $v_l$ and $B_l$ on $l$.  If we assume
that modes with wavelengths larger than $l$ remain the same during the
decay of shorter modes, the configuration of $v$ and $B$ at moment $t$ is
essentially the initial configuration, smeared out over the length scale
$l$.  At the time moment $t$, the value of $v$ and $B$ at a point in space
is roughly the average of the initial $v$ and $B$ over the sphere of
radius $l$ around this point.  Assuming that $v$ and $B$, at $t=0$, is not
correlated over lengths larger than $l_0$ (which is certainly true if
$l_0$ is the Hubble size of the Universe at this time), the average is
over $(l/l_0)^3$ patches with random $v$ and $B$.  Therefore, the order of
magnitude of $v$ and $B$ at the moment $t$ is
\begin{equation}
  v_l \sim v_0 \biggl( {l\over l_0} \biggr)^{-3/2}\qquad
  B_l \sim B_0 \biggl( {l\over l_0} \biggr)^{-3/2}
  \label{vl}
\end{equation}
The characteristic time for the decay of modes with length scale $l$ is
the eddy turnover time,
\begin{equation}
  \tau_l \sim {l \over v_l}
  \label{turb_decay}
\end{equation}
Notice that the decay time (\ref{turb_decay}) does not depend on any
viscosity and resistivity.  This is the consequence of the turbulent
cascade: during the time $\tau_l$ the energy is transfered from modes with
length scale $l$ to $l/2$, then it is cascaded to $l/4$, $l/8$ etc. with
faster and faster rate until the viscosities enter the scene at some small
scale $l_{\text{diss}}$ (which is called the dissipation scale).  Due to
the cascade, the energy (both kinetic and magnetic) is dissipated at a
much faster rate than it would be without turbulence.  The length scale
$l_{\text{diss}}$ does not enter any of our formulae, as we are interested
in the much longer length scales at which the cascade begins; all that
matters is that dissipation occurs at some very small length scale,
terminating the cascade. 

Since $l$ is the smallest wavelength that survives at the moment $t$, its
lifetime must be comparable to $t$.  Solving $\tau_l\sim t$ using Eq.\
(\ref{vl}), one finds that the correlation length scales with time as,
\begin{equation}
  l(t)\sim\left({t\over t_0}\right)^{2/5}l_0
  \label{Saffman} 
\end{equation}
where $t_0=l_0/v_0$ is the eddy turnover time at $t=0$.  From Eqs.\
(\ref{vl},\ref{Saffman}) and the fact that the energy is proportional to
$v^2$ and $B^2$, we find that the latter decays as $t^{-6/5}$.  It is
worth noting that all our arguments leading to these scaling laws are also
valid for decaying hydrodynamic (not MHD)  turbulence.  Indeed, the
$t^{-6/5}$ decay of the turbulent energy is known in fluid dynamics as
Saffman's law \cite{Saffman,Lesieur}. 

The implicit assumption behind our derivation of the $t^{2/5}$ law
(\ref{Saffman}) is that the spectrum at low $k$ remains unmodified during
the decay of high-$k$ modes.  Such an assumption is essential for deriving
Eq.\ (\ref{vl}) (called ``volume averaging'' elsewhere.)  The validity of
this assumption may seem uncertain, since inverse cascade, i.e.\ the
process of transferring energy from high to low modes happens under some
conditions in MHD.  In \cite{Olesen,Olesen-rev}, it was suggested that
inverse cascade is a generic feature of MHD turbulence and could greatly
enhance the correlation length of the magnetic field (although,
coincidentally, the scaling law for the growth of the correlation function
obtained in \cite{Olesen,Olesen-rev} is the same as our Eq.\
(\ref{Saffman}).)  What we want to argue here, to support the derived
scaling laws, is that the inverse cascade of magnetic energy required a
special condition to be met, namely, the average density of the
Chern-Simons number must be non-zero. That the conditions for the inverse
cascade are rather restrictive is a well known fact in plasma physics: 
previous studies \cite{Biskamp,Pouquet,dns} have identified these
conditions to be either a non-zero Chern-Simons number density of the
magnetic field, or a non-helical turbulent flow (see also the discussion
below.)  Since we do not expect the bulk fluid flow to be helical, the
only chance to have an inverse cascade is if the magnetic field has
non-zero Chern-Simons number, which is not the case considered in this
Section.  We have also checked that there is indeed no inverse cascade in
the non-helical MHD turbulence decay by simulating it using the
eddy-damped quasi-normal Markovian (EDQNM) approximation.  Our simulation
will be described in the next subsection. 

\subsection{Numerical simulation of decaying non-helical turbulence in a
non-expanding Universe}

To verify the absence of the inverse cascade, as well as to check our
scaling laws, we use the EDQNM approximation for numerical simualations. 
A full discussion of the merits and shortcomings of the EDQNM
approximation can be found in Refs.\ \cite{Lesieur,Biskamp}, here we will
only note that this approximation is a member in a family of closure
schemes, which are similar to the Hartree--Fock approximation in field
theory.  The EDQNM approximation has been found very successful in
reproducing qualitative features of hydrodynamics and MHD as the
Kolmogorov spectrum of driven hydrodynamic turbulence, the inverse cascade
in MHD (under the above-mentioned conditions), etc.  One should be aware
that it is unclear how the EDQNM approximation can be systematically
improved.  However, it should be adequate to answer our qualitative
questions. 

In the closure schemes, the turbulence is characterized by two functions
$E_k$ and $M_k$, which are the spectra of kinetic and magnetic energy (in
the case of helical flow and helical magnetic fields, one should introduce
two more functions of $k$.) The full set of EDQNM equations are quite
complicated \cite{Pouquet}.  In the Appendix, we reproduce the equations
for the case of non-helical turbulence.  As the initial condition at
$t=0$, we choose the initial kinetic and magnetic spectra to have the same
functional form,
\begin{equation}
  E_k = 0.9 {4\over\pi} {k^2\over(k^2+1)^2},\quad
  M_k = 0.1 {4\over\pi} {k^2\over(k^2+1)^2}
  \label{incon} 
\end{equation}
The $k^2$ behavior of $E_k$ and $M_k$ at low $k$'s comes from the
assumption that the initial state consists of uncorrelated
eddies\footnote{To see that, one could use the following argument.  The
typical wavelength in the initial state is $k_0\sim l_0^{-1}$.  If one
throws away all modes with momentum larger than some $k\ll k_0$, this is
equivalent to smearing out the velocity and the magnetic field over the
distance of $k_0^{-1}$.  As discussed before, if $v$ and $B$ do not
correlate over lengths larger than $k_0^{-1}$, the average takes both down
by a factor of $(k/k_0)^{3/2}$.  The energy, which is proportional to
$v^2$ and $B^2$, is down by a factor of $(k/k_0)^3$. Compare this with the
case of an energy spectrum with $k^2$ behavior at low $k$, in which
throwing away modes with energy larger than $k$ makes the energy of
remaining modes proportional to $\int_0^k\!dk\,k^2\sim k^3$.}. In the
initial condition (\ref{incon}), the magnetic energy constitutes 10\% of
the initial total energy at $t=0$, which is equal to 1.  In choosing the
initial magnetic energy to be small compare to the total energy, we obtain
as a bonus the possibility of checking whether equipartition is reached
during the decaying turbulence.  We choose $\nu=\eta=10^{-3}$, which
correspond to the Reynolds numbers of order $10^3$ at $t=0$.  This is much
smaller than the cosmological Reynolds numbers, but large enough for the
physics at the integral scale to be insensitive of the exact values of
$\text{Re}$ and $\text{Re}_{\text{M}}$.  We use logarithmic discretization
in the $k$ space, with 4 points per octave (we have checked that doubling
the number of points per octave does not alter the result in any
substantial way), and the simplest first-order time-update algorithm. 

\begin{figure}
\begin{center}
\leavevmode
\epsfxsize=5in \epsfbox{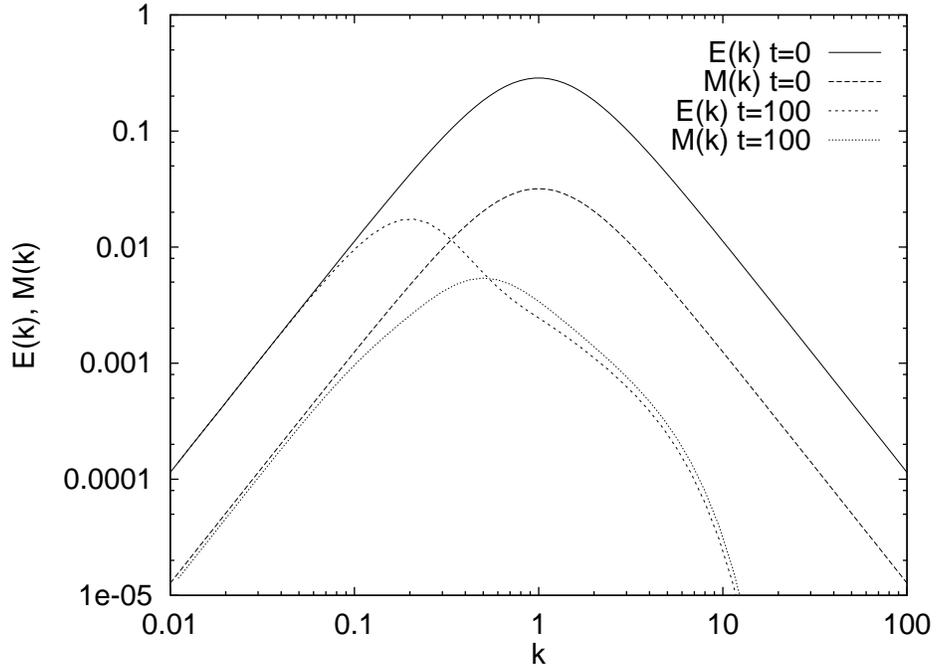} 
\end{center}
\caption{The kinetic and magnetic spectra at $t=0$ and $t=100$}
\label{fig:spectrum}
\end{figure}

\begin{figure}
\begin{center}
\leavevmode
\epsfxsize=5in \epsfbox{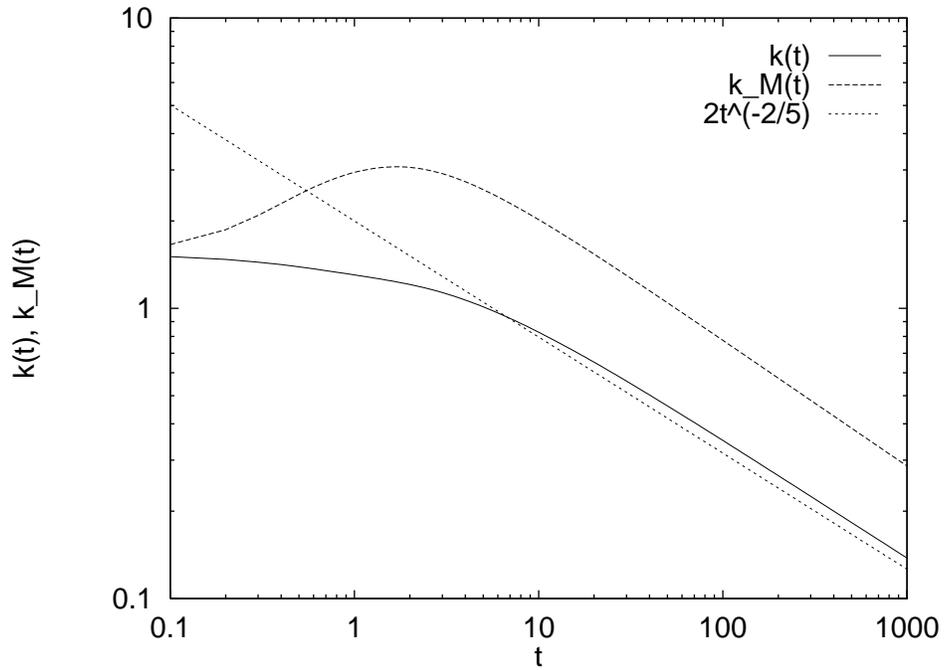}
\end{center}
\caption{Time evolution of the typical wavelength}
\label{fig:Saffman}
\end{figure}

\begin{figure}
\begin{center}
\leavevmode
\epsfxsize=5in \epsfbox{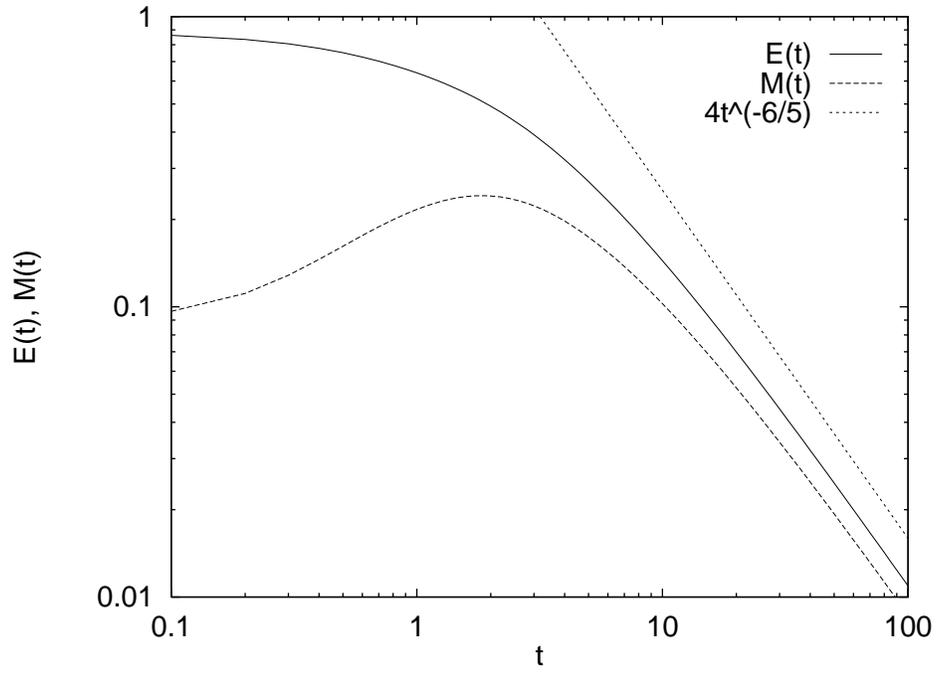}
\end{center}
\caption{Time evolution of the total kinetic and magnetic energy}
\label{fig:E_t}
\end{figure}

The main results are summarized in the Fig. 
(\ref{fig:spectrum}--\ref{fig:E_t}).  In Fig.\ \ref{fig:spectrum}, the
spectra of kinematic and magnetic degrees of freedom are plotted at $t=0$
and $t=100$ (measured in units of the eddy turnover time at $t=0$).  We
see that the low-$k$ part of the spectrum is not modified by turbulent
decay, which means that there is no inverse energy cascade.  It is
interesting to note that, at $t=100$, in the high-$k$ part of the spectrum
the magnetic energy exceeds the kinetic energy, and there is a range of
$k$ where the Iroshnikov--Kraichnan spectrum $E_k\sim M_k\sim k^{-3/2}$ is
observed. Both features have been seen before in the simulation of forced
(not decaying) turbulence \cite{Pouquet}.  In Fig.\ \ref{fig:Saffman},
the typical wavelengths of kinetic and magnetic modes, which are defined
as
\[
  k = {\int\!dp\,E_p \over \int\!dp\,p^{-1}E_p},\qquad
  k_{\text{M}} = {\int\!dp\,M_p \over \int\!dp\,p^{-1}M_p}
\]
are plotted versus time from $t=0$ to 1000.  Both $k(t)$ and
$k_{\text{M}}(t)$ decreases at large $t$ and their time dependence agrees
very well with our prediction, $k\sim t^{-2/5}$.  The ratio
$k_{\text{M}}(t)/k(t)$ approaches a finite value (about 2) at large $t$.
We conclude that our simple arguments reproduce correctly the behavior of
the correlation length in the non-helical turbulence decay.  In Fig.\
\ref{fig:E_t}, we plot the dependence of the total kinetic and magnetic
energy as functions of time. The long-time behavior is well described by
Saffman's law $E\sim t^{-6/5}$. Notice that the magnetic energy grows for
a few eddy turnover periods before decaying.  Apparently, equipartition is
at work here: at first some energy is pumped from the kinetic motion into
the magnetic field. Equipartition is never fully realized, but the ratio
of the magnetic energy to the kinetic energy approaches a value closed to
1 (namely, 0.8) in our simulation.  Therefore, we have seen that our
qualitative arguments (the most important being the lack of inverse
cascade) agree well with the numerical simulation. 

\subsection{Expanding Universe}

So far we have considered the case of a non-expanding Universe.  In the
radiation-dominated Universe, the correlation length, in comoving
coordinates, grows as $\tau^{2/5}$, where $\tau$ is the conformal time. 
Since $\tau\sim T^{-1}$ ($T$ being the temperature), we conclude that in
this case the growth of the correlation length follows the $T^{-2/5}$ law
(not counting the trivial redshift). 

In the real situation we have a hybrid of the two cases.  If at $t=0$ the
eddy turnover time $t_0\sim l_0/v_0$ is smaller than the age of the
Universe, there will be a substantial period of time when the expansion of
the Universe can be neglected, that is, when $0<t\alt t_H$. The expansion
factor for the length is then $(t_Hv_0/l_0)^{2/5}$.  When $t\agt t_H$, one
has to take into account the expansion, and the increase in the
correlation length (not counting the expansion) is
$(T_{\text{EW}}/T_0)^{2/5}$, where $T_0$ is the temperature of the
Universe when the fluid flow becomes non-turbulent (we determine $T_0$
below.)  Therefore, the final correlation length at the present epoch is,
\begin{equation}
  l_{\text{now}} = l_0 \cdot \biggl({t_Hv_0\over l_0}\biggr)^{2/5}
  \cdot \biggl({T_{\text{EW}} \over T_0}\biggr)^{2/5}\cdot
  {T_{\text{EW}} \over T_{\text{now}}} \, ,
  \label{lnow}
\end{equation}
where the last factor comes from the expansion of the Universe.  Putting
in numbers, using for $T_0$ the temperature of $e^+e^-$ annihilation
(which is of order $m_e$), using $v_0\sim0.1$ and $l_0\sim 10^{-2} a_H$,
one finds $l_{\text{now}}\sim 100$ AU.  This scale is larger than the
scale one would have if the only mechanism of dissipation of magnetic
energy is resistive diffusion (about 1 AU \cite{Olinto}), but it is still
too small to be a direct seed for the mean-field dynamo.  The situation is
better if one takes larger values of $v_0$ and $l_0$. However, even in the
extreme case when the magnetic field is initially correlated on the
horizon size and its initial energy density is comparable to that of the
whole Universe, the correlation length is still less than $10^4$ AU. 

Let us check that $T_0$ is the temperature at which $e^+e^-$ annihilation
occurs, which is to say that below this temperature, turbulence ceases,
and therefore so does the enhancement of the magnetic correlation length. 
To this end, we need to compute the Reynolds number $\text{Re}$ before and
after the $e^+e^-$ annihilation.  At $T_{e^+e^-}\sim 1$ MeV, using Eq.\
(\ref{lnow}) and replacing $T_{\text{now}}$ by $T_{e^+e^-}$, one finds the
correlation length $l\sim 10^{19}~\text{GeV}^{-1}$.  The velocity at $T_0$
is
\[
  v \sim v_0 \cdot \biggl({t_Hv_0\over l_0}\biggr)^{-3/5}
  \cdot \biggl({T_{\text{EW}} \over T_0}\biggr)^{-3/5} \sim 10^{-5}
\]
and the viscosity is $\nu\sim (\alpha^2T)^{-1}\sim 10^7\text{GeV}^{-1}$. 
The kinematic Reynolds number is then $10^5$. 

After $e^+e^-$ annihilation, however, the cosmic fluid becomes much more
viscous.  This is the result of the dramatic drop of the density of
charged particles, which leads to a very long mean free path for photons,
which now dominate the viscosity.  The viscosity jumps by the
photon-to-baryon ratio, $10^{10}$, and the Reynolds number becomes small. 
Therefore, $e^+e^-$ annihilation is the moment when turbulence
terminates\footnote{It is sometimes, based on the results of
\cite{jedamzik}, argued that the large diffusion length of photons leads
to damping of the magnetic field, similar to the Silk damping of density
fluctuations.  However, Ref.\ \cite{jedamzik} deals with the damping of
{\em MHD waves} but not the background magnetic field on top of which
these waves propagate (see also the analysis in Ref.\
\cite{Subramanian}.)}.  After this moment, the magnetic field is frozen
into the essentially non-moving fluid and the correlation length increases
only by the expansion of the Universe, as we assumed in deriving Eq.\
(\ref{lnow}).

\section{Decay of helical turbulence}

\label{sec:hel}

Let us now turn to the case when the original magnetic field is helical. 
Here helicity refers to the Abelian Chern-Simons number,
$N_{\text{CS}}=\int\!d{\bf x}\,{\bf A}\cdot{\bf B}$, which is called the
magnetic helicity in plasma physics.  The situation when the initial field
has a non-vanishing Chern-Simons number is not entirely hypothetical,
since some mechanisms indeed generate such helical fields
\cite{Shaposhnikov}.  However, our interest will be mostly in finding out
the largest length scale one could possibly get from a magnetic field from
the electroweak epoch.

It is a well known fact in plasma physics that the Chern-Simons number is
an approximately conserved quantity of MHD turbulence \cite{Biskamp}.  If
the fluid is an ideal conductor, the conservation of $N_{\text{CS}}$ is
exact. This can be seen from the geometrical interpretation of the
Chern-Simons number: $N_{\text{CS}}$ is proportional to the number of
links of the magnetic field lines. When $\eta=0$, the latter are frozen in
the moving fluid, and the number of links cannot change with time.  At
nonzero $\eta$, the conservation of $N_{\text{CS}}$ is less obvious, since
another conserved quantity, the energy $E={1\over2}\int\!d{\bf
x}\,((\rho+p)v^2+B^2)$, dissipates at a finite rate at arbitrarily small
$\nu$ and $\eta$.  The smallness of the dissipation of $N_{\text{CS}}$ can
be explained in the following (non-rigorous) way, which relies only on the
difference between the dimensions of $E$ and $N_{\text{CS}}$.  Let us
write down the equations for the dissipation of the energy and the
Chern-Simons number, which can be derived from Eq.\ (\ref{MHD})
\cite{Biskamp},
\begin{eqnarray}
  \dot{E} & = & -\int\!d{\bf x}\,
  (\nu(\rho+p)(\mbox{\boldmath$\nabla$}\times{\bf v})^2+
  \eta(\mbox{\boldmath$\nabla$}\times{\bf B})^2)
  \label{dE/dt}\\
  \dot{N}_{\text{CS}} & = & -2\eta\int\!d{\bf x}\,{\bf B}\cdot
  (\mbox{\boldmath$\nabla$}\times{\bf B})
  \label{dH/dt} 
\end{eqnarray}

Let us compare the order of magnitude of the right hand sides of Eqs.\
(\ref{dE/dt},\ref{dH/dt}), assuming for simplicity $\nu\sim\eta$ and
equipartition, $\rho v^2\sim B^2$.  If both $E$ and $N_{\text{CS}}$ are
dominated by the largest length scale of turbulence (the integral scale)
$l$, then $N_{\text{CS}}/E\sim l$.  On the other hand, the dissipation of
both energy and the Chern-Simons number occurs at the dissipation scale
$l_{\text{diss}}$.  Since $l_{\text{diss}}$ is much smaller than $l$ (by a
power of the Reynolds numbers), we conclude that
$\dot{N}_{\text{CS}}/\dot{E}\ll N_{\text{CS}}/E$, which means that the
dissipation rate of the Chern-Simons number goes to zero in the limit of
small viscosity and conductivity, in contrast to the energy which
dissipates with a finite rate in this limit.  When the Reynolds numbers
are large, one can neglect the dissipation of the Chern-Simons number (but
not of the energy.)

The conservation of $N_{\text{CS}}$ has an important consequence for the
evolution of the magnetic field.  When $N_{\text{CS}}$ is non-vanishing,
the short-scale modes are not simply washed out during the decay: their
magnetic helicity must be transfered to the long-scale ones.  Along with
the magnetic helicity, some magnetic energy is also saved from turbulent
decay.  This process is known as the inverse cascade
\cite{Biskamp,Pouquet}\footnote{Inverse cascade occurs also in
two-dimensional hydrodynamics and MHD.  In both cases, the conserved
quantities governing the inverse cascade, the enstrophy and the mean
square magnetic potential \cite{Biskamp}, are positively defined, so the
inverse cascade does not requires special initial conditions.}.  The
coupling with kinetic motion ensures that a part of the energy cascaded to
large scales also goes to the kinetic energy of fluid motion, maintaining
equipartition.  We will assume maximal helicity, i.e. ${\bf A}\cdot{\bf
B}$ has the same sign in all space, so the magnetic helicity density is
proportional to $AB\sim lB^2$. Since this quantity is conserved during the
decay of the turbulence, the field strength should scale with the
correlation length as
\[
  B_l\sim B_0\left({l\over l_0}\right)^{-1/2}
\]
This law replaces Eq.\ (\ref{vl}).  This corresponds to ``line averaging''
that gives a much larger amplitude of the magnetic field than the ``volume
averaging \cite{Dimopoulos}.  However, our averaging is not postulated,
but is a dynamical effect which occurs only in the helical case.  An
argument similar to the one used in the non-helical case now gives the
following decay law for helical MHD turbulence: 
\[
  l\sim l_0\left({t\over t_0}\right)^{2/3}\ ,
\]
which replaces the formula (\ref{Saffman}). Therefore, the correlation
length grows faster if the initial field configuration is helical. 
Obviously, this comes from a slower decay of the magnetic field, which
implies a larger equipartition velocity and hence a shorter relaxation
time compared to the non-helical case.

To see what is the largest magnetic length one can get, let us assume that
at $T=100$ GeV one has $l_0\sim a_H$ and $v_0\sim 1$.  Assuming the
turbulence has all decayed at $T_0$, the correlation length at $T=T_0$ is
\[
  l = l_0 \biggl( {T_{\text{EW}} \over T_0} \biggr)^{5/3}
\]
The velocity now scales as $v\sim (T_{\text{EW}}/T_0)^{-1/3}$.  In the
helical case, with our choice of parameters, the turbulence in fact
survives beyond $e^+e^-$ annihilation.  When $T\ll m_e$, the viscosity of
the Universe is mostly due to photons and is given by
\[
  \nu \sim {1\over n_e \sigma_T} \sim {m_e^2 \over \eta_B \alpha T^3}
\]
where $n_e$ is the electron density, $\eta_B$ is the baryon-to-photon
ratio, and $\sigma_T \sim \alpha m_e^{-2}$ is the Thompson cross section. 
One sees that the kinematic Reynolds number drops below 1 at $T_0\sim 100$
eV.  Therefore, the correlation length is now
\[
  l \sim l_0 \cdot \biggl( {T_{\text{EW}} \over T_0} \biggr)^{5/3}
             \cdot \biggl( {T_0\over T_{\text{now}}} \biggr) 
  \sim 100~\text{pc}
\]
Since the collapse of proto-galaxies will reduce this correlation length
by a factor of 100, the galactic seed field will be correlated on a scale
of order 1 pc, still much smaller than the scale of the turbulent motion
of the interstellar gas which we recall is of order 100 pc.  Therefore,
the presence of helicity in the initial magnetic field strongly enhances
the magnetic correlation length, but still cannot make it large enough to
be the seed for the mean-field galactic dynamo. 

\section{Conclusion}

\label{sec:concl}

In this paper we have considered the behavior of the correlation length of
the primordial magnetic field, if such a field is generated in the early
Universe.  We found the decay law of the MHD turbulence, which is
responsible for a substantial increase of the correlation length.  We also
observed a qualitative difference between the cases of non-helical and
helical initial magnetic field.  Taking the case in which the magnetic
field is generated during the electroweak phase transition as an example,
we find that if the magnetic field is not helical the factor gained from
MHD is about $10^2$ and today the field may be correlated at scales as
large as 100 AU.  If by some chance the field is helical, the enhancement
factor is much larger, but the final correlation length cannot exceed 100
pc, corresponding to 1 pc after proto-galactic collapse. 

Let us note some uncertainties remaining in our estimation.  (In general,
addressing these issues would lead to less optimistic values of the
enhancement factor.)  First, it is not at all clear whether in MHD
turbulence the magnetic energy reaches equipartition with the kinetic
energy of bulk fluid motion.  While semi-analytical calculations
(including our simulation) favor equipartition \cite{Pouquet}, some
numerical results indicate that the mean magnetic energy density remains
small and concentrated at scales shorter than the largest turbulence scale
\cite{dns}.  If the latter remains true at very high Reynolds number, the
magnetic field will be smaller after the electroweak phase transition and
the correlation length at present will be smaller than we have estimated. 

The second factor is neutrino diffusion.  After the electroweak phase
transition, neutrinos are the particles with the longest mean free path.
There is a certain time interval when neutrinos are still not decoupled
from the physics at the turbulence scale, but their diffusion length is so
large that the neutrino contribution to viscosity makes the Reynolds
number to drop below 1.  During this time, there is no turbulence and the
magnetic field is frozen.  After the neutrinos decouple from the fluid
motion at the turbulence scale, magnetic stress leads to restoration of
turbulence, and the magnetic length may continue to grow.  The overall
effect is some reduction in the estimate of the final magnetic correlation
length. 

It would be nice to end this paper on a more positive note, and to this
end we turn now to investigating the QCD phase transition, which occurs
later than the electroweak transition and so may yield a longer
correlation length.  If the QCD phase transition is first order, it will
introduce fresh turbulence stronger than the decaying electroweak
turbulence.  The amplitude and correlation length of the magnetic field
will be determined by the QCD turbulence.  If fields were regenerated at
the QCD transition on length scales of order of the horizon, the length
scale it would have today is very large, since the horizon size at the QCD
phase transition is much larger than at the electroweak epoch.  Even with
no enhancement from MHD effects, the expansion of the universe yields a
magnetic correlation length of order 1 pc. It can be estimated that the
turbulence will survive till at least matter-radiation decoupling, at
which the correlation length is is of order 1 kpc in the non-helical case
and 100 kpc if the field is helical.  However, a magnetic field correlated
on the horizon size is not expected to be produced at the QCD phase
transition, and these estimates are too optimistic. Instead, the initial
length scale must be of the order of the bubble spacing at the end of the
phase transition (which is the natural scale of turbulence).  Bubble
spacings larger than 100 cm are unlikely \cite{Ignatius} and may have
problem with standard big bang nucleosynthesis \cite{bbn}.  A bubble
spacing smaller than 100 cm again implies a very small correlation length
at the present epoch.  If this constraint is respected, our analysis of
the QCD case must end pessimistically, as in the electroweak case. 
However, one cannot completely rule out the possibility of a very large
bubble spacing, which leads to a magnetic field correlated on a long
enough length scale to be of interest.  A nonstandard evolution in which
mixing after the phase transition occurs rapidly by hydrodynamic flows
instead of slowly by diffusion or nonstandard nucleosynthesis may be
required in this case.

Finally, although in this paper we presented the smallness of the magnetic
correlation length as undesirable and tried to overcome it by invoking MHD
turbulence, there may exist a non-standard dynamo mechanism where a
small-scale seed gives rise to a large-scale magnetic field (see, e.g.,
\cite{Chandran}).  This possibility is, however, outside the scope of the
present paper. 

\acknowledgments

The author thanks R.~Jackiw, S.-Y.~Pi, E.~Vishniac, and especially
J.~Berges and K.~Rajagopal for stimulating discussions.  He thanks
H.~Kurki-Suonio and K.~Subramanian for comments on the manuscript.  This
work is supported in part by funds provided by the U.S.  Department of
Energy (D.O.E.) under cooperative research agreement \#DF-FC02-94ER40818.

\appendix

\section{The EDQNM approximation}

As mentioned in the main text, in closure schemes, the non-helical
turbulence is characterized by the kinetic and magnetic energy spectra,
$E_k$ and $M_k$, defined so that the total kinetic and magnetic energies
are $\int_0^\infty\!dk\,E_k$ and $\int_0^\infty\!dk\,M_k$, respectively
(we assume isotropy, so the spectra depend only on the absolute value of
${\bf k}$.)  The evolution of the spectra, in the EDQNM approximation, are
described by the following equations,
\begin{eqnarray}
  ( \partial_t + 2 \nu k^2) E_k & = &
  \int\!dp\,dq\, \theta_{kpq} \biggl[
  {1\over q}(xy+z^3) (k^2 E_p E_q - p^2 E_q E_k)
  -{p^2\over q} z(1-y^2) M_q E_k + \nonumber\\
  & & + {k^2\over q} z(1-y^2) M_p M_q \biggr] \nonumber\\
  (\partial_t + 2 \eta k^2) M_k & = &
  \int\!dp\,dq\, \theta_{kpq} \biggl[
  {k^3\over pq} (1+xyz) M_p E_q - {kp\over q} (1-y^2) E_q M_k -
  \nonumber\\
  & & - {k^2\over q} z(1-y^2) M_q M_k \biggr]
  \label{EDQNM}
\end{eqnarray}
which are the non-helical version of the full set of EDQNM equations that
can be found in Ref.\ \cite{Pouquet}.  In Eq.\ (\ref{EDQNM}), the
integration over $p$ and $q$ is perform in the region where $k$, $p$ and
$q$ form a triangle; $x$, $y$ and $z$ are the cosines of the angles of
this triangle,
\[
  x={p^2+q^2-k^2\over2pq},\quad y={q^2+k^2-p^2\over2qk},\quad
  z={k^2+p^2-q^2\over2kp}
\]
and $\theta_{kpq}$ is defined as
\[
  \theta_{kpq} = {1 - \text{e}^{-\mu_{kpq} t}\over\mu_{kpq}},\qquad
  \mu_{kpq} = \mu_k + \mu_p + \mu_q
\]
where
\[
  \mu_{k} = (\nu+\eta)k^2 + C_s \biggl( \int_0^k\!dq\,
  q^2(E_q+M_q)\biggr)^{1/2} + 
  C_a k\biggl( \int_0^k\!dq\, M_q\biggr)^{1/2}
\]
$C_s$ and $C_a$ are phenomenological constants, the exact values of which
are not qualitatively important.  Following \cite{Pouquet}, we choose
$C_s=0.26$ and $C_a=1/\sqrt{3}$.

\end{document}